\documentclass[a4paper,twocolumn,superscriptaddress,prl, aps, floatfix ]{revtex4-1}

\usepackage{epsfig}
\usepackage{verbatim}	 
\usepackage{amsmath,amssymb}
\usepackage{calrsfs}
\usepackage{times}
\usepackage{amsmath}
\usepackage{psfrag}
\usepackage{color}
\usepackage[latin1]{inputenc}
\usepackage[english]{babel}
\usepackage{graphicx}
\usepackage{braket}
\usepackage{natbib}
\usepackage{hyperref}
\usepackage{lipsum}
\usepackage{multirow}
\usepackage{siunitx}          
\usepackage{cleveref}
\usepackage[colorinlistoftodos]{todonotes}

\addto\captionsenglish{}

\DeclareMathAlphabet{\mathcal}{OMS}{cmsy}{m}{n}

\newcommand{\rev}[1]{{{#1}}}

\begin{document}
\title{Link prediction in multiplex networks via triadic closure}

\author{Alberto Aleta}
\affiliation{ISI  Foundation,  via  Chisola  5,  10126  Torino, Italy}

\author{Marta Tuninetti}
\affiliation{ISI  Foundation,  via  Chisola  5,  10126  Torino, Italy}

\author{Daniela Paolotti}
\affiliation{ISI  Foundation,  via  Chisola  5,  10126  Torino, Italy}

\author{Yamir Moreno}
\affiliation{Institute for Biocomputation and Physics of Complex Systems (BIFI), University of Zaragoza, 50018 Zaragoza, Spain}
\affiliation{Department of Theoretical Physics, University of Zaragoza, 50018 Zaragoza, Spain}
\affiliation{ISI  Foundation,  via  Chisola  5,  10126  Torino, Italy}

\author{Michele Starnini}
\thanks{Corresponding author: michele.starnini@gmail.com}
\affiliation{ISI  Foundation,  via  Chisola  5,  10126  Torino, Italy}

\begin{abstract}
Link prediction algorithms can help to understand the structure and dynamics of complex systems,
to reconstruct networks from incomplete  data  sets  and  to  forecast  future  interactions  in  evolving networks. 
Available algorithms based on similarity between nodes are bounded by the limited amount of links present in these networks. 
In this work, we reduce this latter intrinsic limitation and \rev{show that different kind of relational data can be exploited to improve the prediction of new links.
To this aim, we propose a novel link prediction algorithm 
by generalizing the Adamic-Adar method to multiplex networks composed by an arbitrary number of layers, that encode diverse forms of interactions. 
We show that the new metric outperforms the classical single-layered Adamic-Adar score and other state-of-the-art methods,  across several social,  biological  and  technological  systems.
As a byproduct, the coefficients that maximize the Multiplex Adamic-Adar metric indicate how the information structured in a multiplex network can be optimized for the link prediction task, revealing which layers are redundant.  
Interestingly, this effect can be asymmetric with respect to predictions in different layers.}
Our work paves the way for a deeper understanding of the role of different relational data in predicting new interactions and provides a new algorithm for link prediction in multiplex networks that can be applied to a plethora of systems. 
\end{abstract}

\maketitle

\rev{Network science has established as a pivotal tool to characterize the structure of real world complex systems, that involves multiple types of relations among their fundamental components \cite{newman10-1}.
One of the most important challenges within the complex systems framework is to elucidate which entities are related to which others and what are the types of these relationships \cite{NIPS2003_2465}.
Within the network science domain, this scientific task translates into a link prediction problem, that attempts to estimate the likelihood of the existence of a link between two nodes, based on the observed links and attributes of nodes \cite{lp_survey, 10.1145/1835804.1835837}. 
Link prediction algorithms are extremely helpful in at least two directions:
to reconstruct networks from incomplete data sets and to forecast future interactions in evolving networks. 
Examples of the first application can be found in biological networks, such as protein interaction networks, where many links are still unknown and their existence must be demonstrated by expensive experiments \cite{Amaral6}. 
Prediction algorithms help in focusing experimental efforts toward those links most likely to exist. 
The second task, link forecasting, is routinely applied in online social networks, such as Facebook. 
New friendship are indeed recommended based on link prediction algorithms, so that individuals can efficiently find peers they are interested in \cite{Nowell2003, Hasan2011}.}

The link prediction problem is a long-standing challenge at the intersection between  computer science and statistical physics communities. 
Traditional algorithms include Markov chains and statistical models \cite{10.5555/2073876.2073934}, while recent approaches from the physics community, such as random walk processes and
maximum likelihood methods,  have been considered \cite{lp_rw, Guimera22073}. 
Link prediction algorithms can be classified mainly into two categories: similarity based methods and probabilistic models \cite{Marjan2018Oct,MartnnezVnctor2016Dec}. 
Since the latter can be computationally unfeasible for large networks, a lot of attention has been devoted to the creation of good similarity scores. Many of these similarity methods are based on the same basic idea, two nodes are likely to be linked if they share a common neighbor \cite{Bliss2014Sep, Adamic2003Jul}. 
Despite its simplicity, this concept has proven to be quite useful for highly assortative networks, such as scientific collaboration networks \cite{Clauset2008May}. 
However, as signaled by Jia et al. \cite{Jia2020}, the prediction power of any similarity-based link prediction algorithm is bounded due to the limited amount of links present in the network.

\rev{In network science, richly structured data can be represented by multilayered networks, in which each layer accounts for a different type of interaction \cite{Kivela2014Sep,Aleta2019Mar}.
For instance, social interactions can have different purposes (e.g., leisure vs work) and happen through various communication channels, including face-to-face interactions, e-mail, Facebook, phone calls and so on.}
\rev{The idea of predicting links in multilayer networks has been explored during the last decade from several different points of view. 
For instance, Davis et al. \cite{Davis2013Jun} proposed a technique to include multi-relational data for link prediction from a probabilistic point of view. Similarly, several extensions of probabilistic models to multilayer networks have been proposed \cite{Kleineberg2016Jul,Hassan2006,Lu2010Dec,Matsuno2018,Pujari2015,DBLP:conf/allerton/SotiropoulosBPT19}. 
Other works, rather than focusing on incorporating new data to already existing networks, used multilayer structures to focus on the temporal evolution of the networks \cite{Hajibagheri2016Aug,Yao2016Jan}. 
Several studies extended the notion of neighborhood to multilayer networks \cite{Jalili2016Nov,Hristova2016Dec,Mandal2018Nov,Junuthula2018}, focusing on networks of two layers. }
\rev{However, a fundamental question is still unanswered: How can different kind of relational data be exploited to improve the prediction of new interactions?
For instance, to which extent are face-to-face interactions predictive of new Facebook friendship?
Interestingly, it has been recently showed that the multiplex network representation can be redundant in some cases, as the information encoded in some layers can be effectively included in others, reducing the number of layers \cite{str_red}.
Therefore, how can link prediction algorithms optimize the information structured in a multiplex network representation, that can be sub-optimally organized?}

\rev{In this paper, we address these questions by proposing a novel metric for link prediction in multiplex networks, based on a generalization of the Adamic-Adar method for single-layered networks \cite{Adamic2003Jul}. 
Our metric fully exploits the complexity of the relationships that might be established across the fundamental components of complex systems, by considering all possible triadic closures in the corresponding multiplex representation. 
We show that this score, that can be applied to any multiplex topology composed by an arbitrary number of layers, is able to outperform other metrics based on single-layered similarity between nodes, across several social and biological systems. 
We show that the information encoded in different layers can be asymmetric with respect to the link prediction problem: e.g., face-to-face interactions can be partially predictive of new Facebook friendship, but not vice versa.
}



\rev{
We consider 8 different data sets spanning several types of social, biological and technological systems, represented as multiplex networks.
i) Copenhagen Networks Study (CNS), 4 layers represent physical proximity, phone calls, text messages, and Facebook friendships among university students \cite{Sapiezynski:2019aa};
ii) C. Elegans genetic (CEG): genetic and protein interactions of the C. Elegans, 3 layers represent direct, physical, and additive genetic interactions \cite{Stark2006Jan}; 
iii) C. Elegans neural (CEN): neural network of the C. Elegans, 3 layers represent electric, chemical monoadic, and chemical polyadic interactions \cite{Chen2006Mar}; 
iv) CS-Aarhus (CSA): social network of employees of the Computer Science department at Aarhus, 5 layers represent Facebook, leisure, work, co-authorship, and lunch interactions \cite{Magnani2013Mar};
v) CKM Physicians (CKM): a social network of physicians, 3 layers represent who they ask for advice, who they discuss cases with, and who are their friends \cite{Coleman1957Dec}; 
 vi) EU air (EUA): air transportation network of Europe, 27 layers represent airlines routes \cite{Cardillo2013Feb}; 
vii) Lazega (LAZ): social network of partners and associates of a corporate law partnership, 3 layers represent co-work, friendship and advice \cite{Lazega}; 
viii) Vickers (VIC): social network of students in a school in Victoria, Australia, 3 layers represent who they get on with, who are their best friends and who they prefer to work with \cite{Vickers}.  See Table 1 of the Supplementary Material (SM) \cite{SM} for details about the data sets.}

In the following, we will contrast different algorithms for link prediction on these data sets. The quality of link prediction algorithms can be evaluated by two metrics: the Receiver Operating Characteristics (ROC) curve, with the corresponding Area Under the Curve (AUC) value, and the Precision. The Precision can be computed as $n^\ast/n$, where $n$ is the number of new links that we want to predict and $n^\ast$ is the amount of correct predictions among the top $n$ links.  Thus, it provides complementary information to the one given by the AUC.
It is important to highlight that, due to the limited amount of links present in a network, the AUC of any similarity-based link prediction algorithm is bounded \cite{Jia2020}. For instance, if similarity is based on common neighbors, two nodes without any neighbor in common will have a score equal to zero. The number of scoreless links bounds the maximum and minimum values of the AUC to $\text{AUC}_\text{min} = \frac{1}{2}(1+p_1)(1-p_2)$ and $\text{AUC}_\text{max} = \text{AUC}_\text{min} + p_1 p_2$, where $p_1$ ($p_2$) is the fraction of links with a score different from 0 among those links that will (will not) exist in the future, see Section 2 of the SM \cite{SM} for details. Note that only when $p_1 = p_2 = 1$, i.e., there are no scoreless links, it holds $\text{AUC}_\text{min} = 0$ and $\text{AUC}_\text{max} =1$.

We propose a generalization of the Adamic and Adar  ($AA$) score  \cite{Adamic2003Jul}, one of the most common and successful methods for link prediction in social networks. 
The $AA$ score between nodes $u$ and $v$ is given by the number of common neighbors weighted by their degree, 
\begin{equation}\label{eq:AA}
AA(u,v) = \sum_{w\in \Gamma(u) \cap \Gamma(v)} \frac{1}{\ln (k_w) }\,.
\end{equation}
where $\Gamma(u)$ represents the set of neighbors of node $u$ and $k_w = |\Gamma(w)|$ is the degree of node $w$.
In a multiplex network, the $AA$ score can be applied to different layers, depending on which layer $\alpha$ the set of neighbors $w \in \Gamma_{\alpha}(u) \cap \Gamma_{\alpha}(v)$ is considered, where $\Gamma_{\alpha}(u)$ represents the set of neighbors of node $u$ in layer $\alpha$.

\rev{For example, let us consider that we are interested in predicting future phone calls among the participants in the CNS.  
The classic $AA$ method considers the set of neighbors in the same phone calls layer.
If the $AA$ score is applied to the layer representing Facebook friendship, instead, the rationale is that two individuals are more likely to interact offline (phone call) if they share many friends on Facebook  (i.e., common neighbors in the Facebook layer). 
The same reasoning applies to other layers. }
Table \ref{tab:tableAA} shows the Precision and AUC values (together with its theoretical bounds) to predict phone calls, obtained for the AA method applied to each layer of the CNS (excluding physical proximity interactions for being much denser than others).
\rev{Interestingly, while the maximum Precision (0.04) is obtained by applying the AA score to the same {calls} layer, the maximum AUC (0.69) is obtained by considering the {Facebook} layer. 
This implies that this kind of interactions (Facebook friendship) include useful information to predict new links not encoded in the {phone calls} layer.
This is also reflected in a larger maximum theoretical bound of the AUC for the {Facebook} layer with respect to the {phone calls} layer.
Note also that by using the aggregated network, in which all layers are projected onto a single one, one obtains maximum AUC but zero Precision.}

\begin{table}[]
\begin{tabular}{@{}llll@{}}
\toprule
\textbf{Method}             & \textbf{Precision} & \textbf{AUC}  & \textbf{AUC {[}worst-best{]}} \\ \colrule
Random                      & 0                  & 0.50          & {[}0-1{]}                     \\
AA${}_\text{calls}$      & \textbf{0.04}                  & 0.60          & {[}0.60-0.60{]}                  \\
AA${}_\text{facebook}$       & 0                  & \textbf{0.69}          & {[}0.59-0.70{]}               \\
AA${}_\text{sms}$          & 0               & 0.60          & {[}0.60-0.60{]}               \\
AA${}_\text{aggregated}$    & 0                  & 0.76          & {[}0.65-0.80{]}                  \\
\botrule
\end{tabular}
\caption{\textbf{Precision and AUC to predict new {phone calls} in the CNS data set}, obtained by using the classical Adamic-Adar metric on each layer (calls, Facebook, texts), and on the aggregated network.
Theoretical bounds of the AUC are showed. 
Predictions are tested over the set of non-overlapping links over all layers (7\% of the total).
Best results among the layers are highlighted in bold.}
\label{tab:tableAA}
\end{table}

\begin{figure}[tbp]
\includegraphics[width=\linewidth]{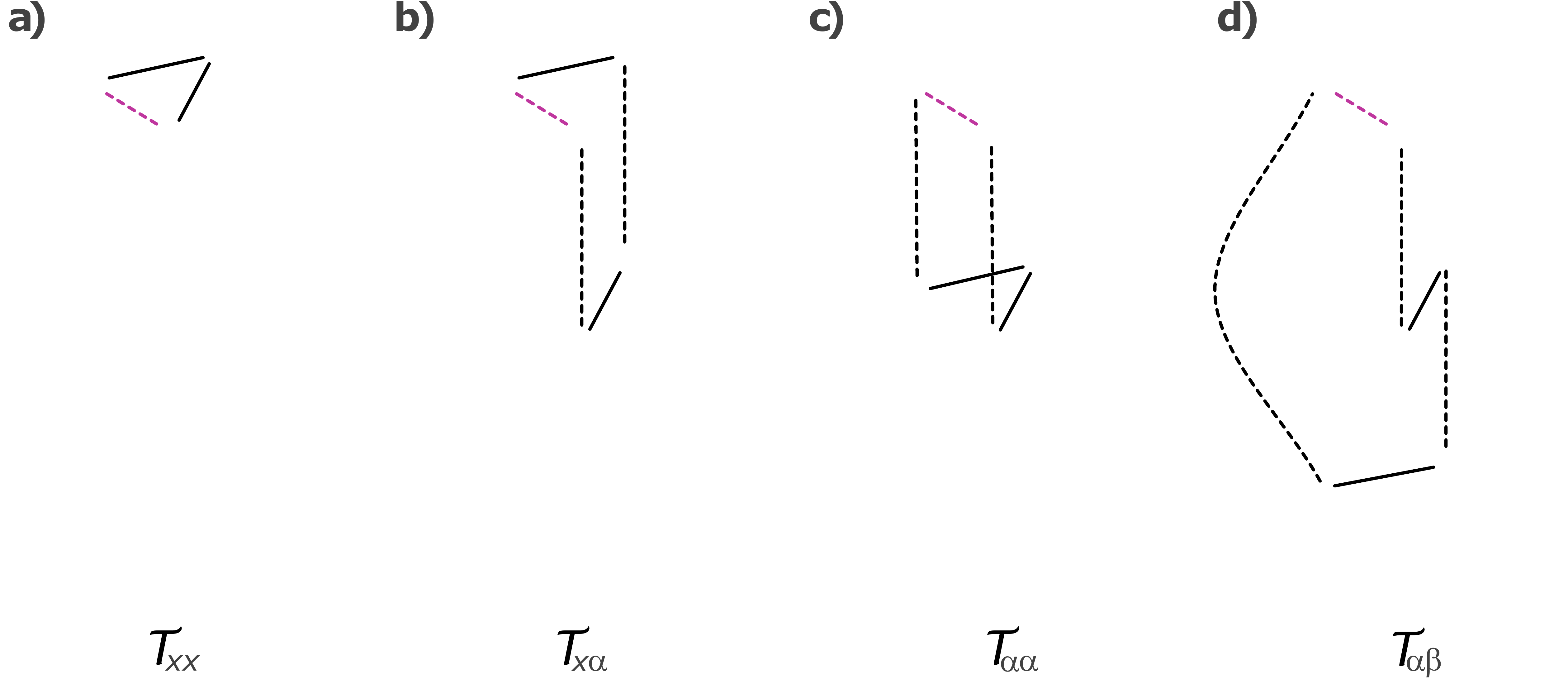}
\caption{\textbf{Triadic relationships in a multiplex network.} Given two nodes $u$ and $v$ for which we want to predict the future existence of a link (red dashed line) in the top layer $x$ (green), based on their connections with another node $w$ (pink) via triadic closure, we can distinguish four types of triadic relationships: (a) $u$ and $v$ are both connected to $w$ in the prediction layer $x$; (b) the link between $u$ and $w$ is in the prediction layer $x$, but $v$ and $w$ are connected in a different layer $\alpha$, or viceversa; 
(c) both $u$ and $v$ are connected to $w$ in a layer $\alpha$ different from the prediction layer $x$; 
(d) $u$ and $w$ are connected in a layer $\alpha$ different from the prediction layer $x$, $v$ and $w$ are connected in a third layer $\beta$ different from layers $\alpha$ and $x$.
}
\label{fig:trian}
\end{figure}

This observation shows the need to go beyond single-layered scores and combine them into a more general metric, that fully exploits the multiplex nature of the networks taken into account. 
Note, indeed, that single-layered metrics considered triadic relations among three nodes $u$, $v$ and $w$, in which 
the two links $u-w$ and $v-w$ lay both in the same layer.
However, triadic relations in multiplex networks can be far more richer \cite{PhysRevE032804, Cozzo2015Jul}. 
Figure \ref{fig:trian} shows different kinds of triadic relations in multiplex networks.
\rev{Let us indicate as $x$ the layer on which the link $u-v$ is to be predicted.}
One can distinguish four types of triadic relations depending on the location of the $(u,w)$ and $(v,w)$ links: 
i) $\mathcal{T}_{xx} = \{(u,v,w) | w \in \Gamma_{x}(u) \cap \Gamma_{x}(v)  \}$, both links lay in layer $x$; 
ii) $\mathcal{T}_{x\alpha} = \{(u,v,w) | w \in \Gamma_{x}(u) \cap \Gamma_{\alpha}(v) \}$ and $\mathcal{T}_{\alpha x}$, one link lays in the layer $x$ and the other lays in another layer $\alpha$;
iii) $\mathcal{T}_{\alpha \alpha} =  \{(u,v,w) | w \in \Gamma_{\alpha}(u) \cap \Gamma_{\alpha}(v)  \}$, both links are in the same layer $\alpha$, different from layer $x$; 
iv) $\mathcal{T}_{\alpha\beta} =  \{(u,v,w) | w \in \Gamma_{\alpha}(u) \cap \Gamma_{\beta}(v) \} $ and $\mathcal{T}_{\beta\alpha}$, 
one link is in layer $\alpha$ and the other in layer $\beta$, both 
different from layer $x$.

Within this formalism, one can consider a score that counts the common neighbors closing triads of each type, and weight each contribution by the logarithm of the degree, as in the Adamic-Adar score,
\begin{equation}\label{eq:MAA}
MAA(u,v)  =  \sum_{\alpha, \beta}    \sum_{w\in\mathcal{T}_{\alpha\beta}}  \frac{\eta_{x\alpha} \eta_{x\beta}}{\sqrt{\langle k \rangle_\alpha \langle k \rangle_\beta}} 		 \frac{1}{\sqrt{\ln(k_w^\alpha ) \ln(k_w^\beta)}}
\end{equation}
This expression is the generalization of the Adamic-Adar score for multiplex networks (MAA) with an arbitrary number of layers, in which the links to be predicted all lay in the same layer $x$. 
Several considerations are in order.

First, the contribution of each triads $(u,v,w) \in \mathcal{T}_{\alpha\beta}$ is weighted by the square root of the logarithm of the degree of node $w$ in the two layers involving $\alpha$ and $\beta$. With this choice, the original weight $1/ \ln(k_w)$ is naturally recovered for $\alpha=\beta=x$. 
Second, note that different layers of a multiplex network may show very different densities, as shown in the Table 1 of the SM \cite{SM}. 
In case of similarity scores based on the number of common neighbors, as in this case, denser layers will have more triads and thus will be less informative. We take into account this by weighting the contribution of each type of triadic relation by the square root of the average degree of the layers involved, $\sqrt{\langle k \rangle_\alpha}$. 
Third, the coefficients $\eta_{x\alpha}$ before each term allow us to control the relative weight of each type of triadic closure in the total score of the link. We choose them in a way that $\eta_{x\alpha}$ corresponds to the weight of layer $\alpha$. 
Without lack of generality, we choose $\sum_{\alpha} \eta_{x\alpha} = 
1$. 
\rev{Fourth, the application of the AA score to layer $\alpha$, corresponding to triads closures  $\mathcal{T}_{\alpha\alpha}$ (case (c) of Fig. \ref{fig:trian}), is recovered by setting $\eta_{x\alpha}=1$. 
The original AA score in single-layered networks (case (a) of Fig. \ref{fig:trian}) is recovered by simply setting $\eta_{xx}=1$.}

\begin{figure*}
\includegraphics[width=\linewidth]{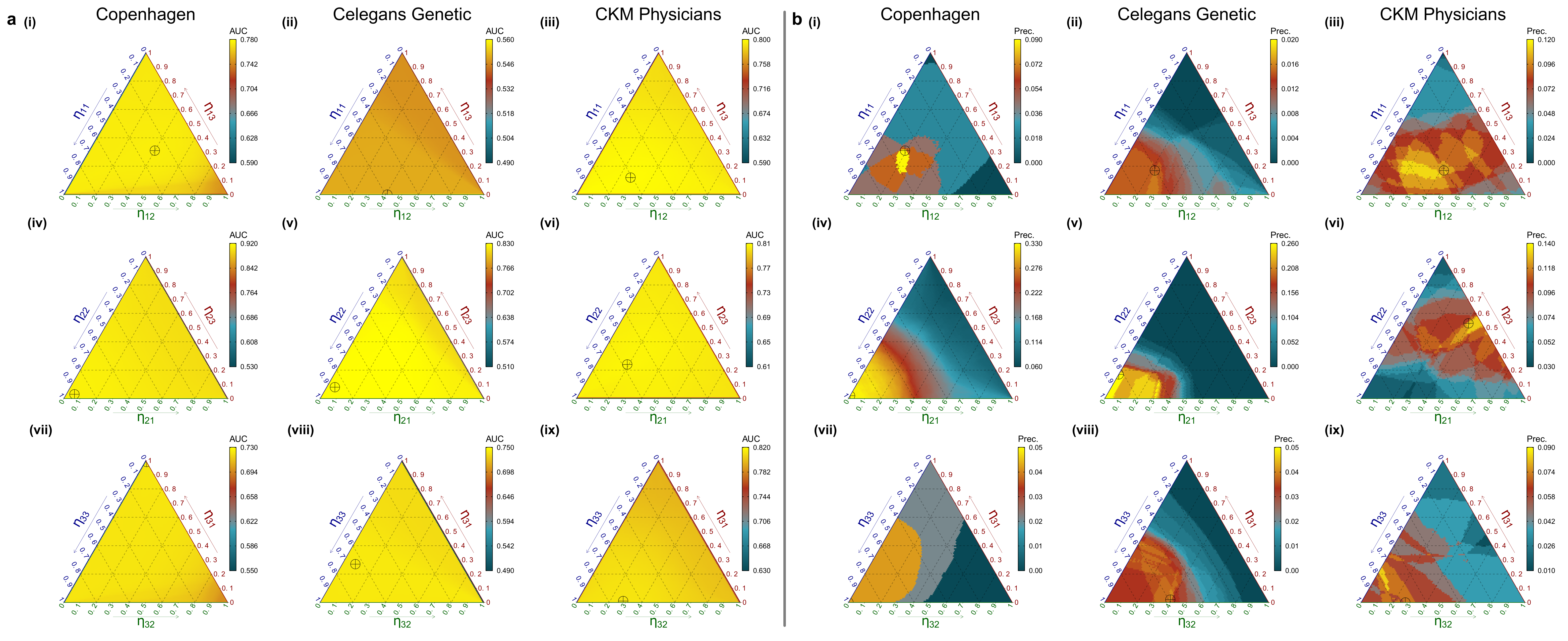}
\caption{\textbf{AUC (panel (a), left) and Precision (panel (b), right) of the MAA metric} for different values of the coefficients $\eta_{x \alpha}$,  indicating the weight of layer $\alpha$ in the prediction of new links in layer $x$.
Varying the values of two coefficients, the third is naturally fixed. 
Each column corresponds to a different data set, represented as a multiplex network of three layers. 
Each row corresponds to a prediction in a different layer $x$, see Table 2 of the SM \cite{SM} for the corresponding interactions.
A cross indicates the maximum value for each plot, corresponding to the combination of coefficients ($\eta_{x1}$, $\eta_{x2}$, $\eta_{x3}$), reported in Table 2 of the SM \cite{SM}, that maximizes AUC or Precision for the prediction of new links in layer $x$. 
\label{fig:coeff}}
\end{figure*}


Figure \ref{fig:coeff} shows the AUC (a) and Precision (b) of the $MAA$ metric as a function of the coefficients $\eta_{x\alpha}$,
\rev{for 3 of the 8 data sets under consideration. Others are showed in the Figures 1 and 2 of the SM \cite{SM}.}
\rev{For the sake of convenient visualization, we consider only three layers for each network, to visualize the three coefficients in a triangle.}
\rev{For each network, we consider prediction of links in each of the three layers.
The coefficient $\eta_{x\alpha}$ indicates the weight of layer $\alpha$ in the prediction of new links in layer $x$. For instance, in the CNS data set, the coefficient $\eta_{12}$ indicates the weight of Facebook friendship (represented in layer 2) in predicting new phone calls (layer 1). 
One can see that, in most cases, the maximum value of the AUC and the Precision is achieved for non-trivial combinations of the coefficients, i.e. different from $\eta_{xx}=1$ which corresponds to the classical AA score, showed in the left corner of triangles. 
This is particularly true for the Precision, whose maximum is achieved in some cases in the middle of the triangle, i.e. with similar contributions for each layer, as in the case of the CKM or CNS networks.}
\rev{The exact values of the coefficients maximizing AUC and Precision for each data set are reported in Table 2 of the SM \cite{SM}.}

\rev{Therefore, Figure \ref{fig:coeff} shows that the prediction of a certain kind of links can be improved by exploiting additional, related information, encoded in other layers.
For instance, Facebook friendship can help in predicting new calls (i.e. the maximum AUC for this task is obtained for $\eta_{12}=0.40$, see plot (i) of panel (a) in Fig. \ref{fig:coeff} and Table 2 of the SM \cite{SM}), or additive genetic interactions and physical association can be predictive of direct protein interactions in C. Elegans (i.e. the maximum Precision is obtained for $\eta_{13}=0.17$ and $\eta_{12}=0.22$, see plot (ii) of panel (b) in Fig. \ref{fig:coeff} and Table 2 of the SM \cite{SM}).} 
\rev{Interestingly, this effect can be asymmetric: new offline interactions (calls and texts) are not predictive of Facebook friendships, as the corresponding coefficients $\eta_{21}$ and  $\eta_{23}$ for this prediction task are zero. This is showed in plots (vii) of panel (a) and panel (b) in Fig. \ref{fig:coeff}: the maximum value of the Precision and AUC is obtained for $\eta_{22} \simeq 1$, in the left corner of the plots (see also Table 2 of the SM \cite{SM}).}
\rev{This implies  that not all layers add valuable information for a specific link prediction task.
In this case, a complete multiplex representation is redundant and such layer can be effectively included in the others without missing relevant information.} 

\rev{Furthermore, we test if the MAA metric is able to optimally extract information from the multiplex representation, compared with the AA score applied to the aggregate network, that include the same amount of information. 
Table \ref{tab:comparison} shows that the MAA metric outperforms the classical AA score with respect to both AUC and Precision,  in all data sets under consideration. 
Finally, in Table \ref{tab:comparison} we compare the MAA metric with other, state-of-the-art metrics for link prediction applied to the aggregated network representation, that includes all information available. In particular: Common Neighbours (CN), Jaccard's Coefficient (JC), and Preferential Attachment (PA), which are based on the one-step neighborhoods of the nodes like the AA score \cite{Gao2015Jun}; and the Katz distance \cite{Katz1953Mar}, which instead is based on path length. Table \ref{tab:comparison} shows prediction of links in the first layer of each data set, predictions in other layers are showed in Table III and IV of the SM \cite{SM}, with similar results. 
One can see that the MAA metric outperforms the Precision of all other metrics in all but one case (the Lazega dataset), 
while it outperforms the AUC of other methods in 5 of 8 data sets under consideration.} 

\begin{table}[]
\begin{tabular}{@{}l|cccccccc@{}}
\toprule
\textbf{Method} & \textbf{CNS} & \textbf{CEG}  & \textbf{CEN}  & \textbf{CSA} & \textbf{CKM} & \textbf{EUA}  & \textbf{VIC}  & \textbf{LAZ}  \\ \colrule
AA    & 0.76           & 0.55          & 0.79          & 0.79          & 0.80           & 0.86          & 0.68         & 0.66 \\
CN    & 0.76           & 0.55          & 0.78          & 0.78          & 0.80           & 0.86          & 0.67          & 0.66 \\
JC    & 0.76           & 0.55          & 0.77          & 0.88          & 0.80           & 0.82          & 0.68          & 0.70 \\
Katz  & 0.72           & 0.59          & 0.77          & 0.74          & \textbf{0.91}  & 0.89          & 0.65          & 0.58 \\
PA    & 0.58           & \textbf{0.65} & 0.62          & 0.39          & 0.64           & \textbf{0.90} & 0.61          & 0.54 \\
MAA   & \textbf{0.77}  & 0.55          & \textbf{0.79} & \textbf{0.91} & 0.80           & 0.87          & \textbf{0.71} & \textbf{0.71} \\
\botrule
AA    & 0.00           & 0.00          & 0.03          & 0.03          & 0.09           & 0.00          & 0.35          & 0.14 \\
CN    & 0.00           & 0.00          & 0.03          & 0.03          & 0.10           & 0.00          & 0.35          & 0.13 \\
JC    & 0.00           & 0.00          & 0.02          & 0.32          & 0.08           & 0.00          & 0.40          & \textbf{0.23} \\
Katz  & 0.00           & 0.00          & 0.00          & 0.00          & 0.00           & 0.00          & 0.31          & 0.00 \\
PA    & 0.00           & 0.00          & 0.02          & 0.00          & 0.01           & 0.02          & 0.27          & 0.09 \\
MAA   & \textbf{0.09}  & \textbf{0.02} & \textbf{0.08} & \textbf{0.39} & \textbf{0.11}  & \textbf{0.19} & \textbf{0.46} & 0.21 \\
\botrule
\end{tabular}
\caption{\textbf{AUC (top) and Precision (bottom) to predict the first layer of each data set}, obtained for different metrics: Adamic Adar (AA), Common Neighbors (CN), Jaccard Similarity (JC), Katz distance (Katz), Preferential Attachment (PA), and Multiplex Adamic Adar (MAA). All metrics except for the MAA are applied to the aggregated network, including all information available. Best method is highlighted in bold. 
}
\label{tab:comparison}
\end{table}

Before concluding, we stress that the metric encoded in Eq. \eqref{eq:MAA} is different from previous extensions of link prediction to multilayer networks. Similarly, other approaches calculate the score of each layer and aggregate all of them (possibly with some weights), effectively neglecting structures of types $\mathcal{T}_{x\alpha}$ and $\mathcal{T}_{\alpha\beta}$ \cite{Sharma2015Nov,Yao2017Jul,Samei2019Oct,DBLP:conf/allerton/SotiropoulosBPT19}. 
To sum up, 
\rev{
we proposed a general method for link prediction that fully exploits different kind of relational data encoded in 
several social and biological networks.}
Our metric is a generalization of the Adamic-Adar score for multiplex networks with an arbitrary number of layers, and it is able to outperform single-layered AA scores \rev{in all considered data sets.
The MAA metric also outperforms several well-known link prediction algorithms, such as the Jaccard's Coefficient or the Katz distance.} 
\rev{The coefficients $\eta_{x\alpha}$ that maximize the MAA score have an interesting interpretation, as they correspond to the weight to be assigned to each layer in order to optimize the information structured in the network for the link prediction task, indicating which layers are redundant. Interestingly, this effect can be asymmetric with respect to predictions in different layers.
The computational complexity of the MAA metric is similar to other similarity-based scores. 
With respect to the classical AA score, it increases with the number of layers in the multiplex network, which is usually small. 
Note that the triadic relationships need to be computed just once and stored, then the whole range of coefficients can be scanned to obtain the ones that maximize the MAA score.}
In future works, it would be interesting to generalize to multiplex networks other metrics based on single layers, such as the Katz distance, which is based on paths that can be reconstructed across layers.

\begin{acknowledgments}
We acknowledge support from Intesa Sanpaolo Innovation Center. Y. M. acknowledges partial support from the Government of Arag\'on and FEDER funds, Spain through grant E36-20R to FENOL, and by MINECO and FEDER funds (grant FIS2017-87519-P). D.P. and M.S. acknowledge financial support from the project Casa nel Parco (POR FESR 14/20 - CANP - Cod. 320 - 16) 
funded by Regione Piemonte.
The funders had no role in study design, data collection, and analysis, decision to publish, or preparation of the manuscript.
\end{acknowledgments}

\bibliographystyle{apsrev4-1}
%

\pagebreak
\newpage
\clearpage
\widetext
\begin{center}
\textbf{\large Supplementary information}
\end{center}

\section{Empirical data sets}

The data sets used in this paper have been collected and analyzed in different studies, cited in the main text. 
In particular, 7 data sets are publicly available in the following repository\footnote{https://comunelab.fbk.eu/data.php}, together with a brief description. 
The remaining data set, the Copenhagen Networks Study,  can be directly downloaded from the reference indicated in the main text.

Table \ref{tab:description} shows the number of nodes and links in each layer for each network under consideration. 
Note that we consider three layers for each network. 
In most cases the density of links greatly varies across layers. 

\begin{table}[h]
\begin{tabular}{@{}llll@{}}
\toprule
\textbf{Network}                           & \textbf{Layer}               & \textbf{Nodes} & \textbf{Links} \\ \hline
\multirow{3}{*}{Copenhagen} & Phone calls                  & 536            & 621            \\ 
                                           & Facebook                     & 800            & 6429           \\
                                           & SMS                          & 568            & 697            \\ \hline
\multirow{3}{*}{C. Elegans genetic}        & Direct interaction           & 3126           & 5472           \\
                                           & Physical association         & 239            & 270            \\
                                           & Additive genetic interaction & 1046           & 2115           \\ \hline
\multirow{3}{*}{C. Elegans neural}         & Electric                     & 253            & 517            \\
                                           & Chemical Monadic             & 260            & 888            \\
                                           & Chemical Plyadic             & 278            & 1703           \\ \hline
\multirow{3}{*}{CS-Aarhus}                 & Facebook                     & 60             & 193            \\
                                           & Leisure                      & 32             & 124            \\
                                           & Lunch                        & 60             & 194            \\ \hline
\multirow{3}{*}{CKM Physicians}            & Advice                       & 215            & 449            \\
                                           & Discussion                   & 231            & 498            \\
                                           & Friendship                   & 228            & 423            \\ \hline
\multirow{3}{*}{EUair}                     & Airline1                     & 106            & 244            \\
                                           & Airline2                     & 128            & 601            \\
                                           & Airline3                     & 99             & 307            \\ \hline
\multirow{3}{*}{Lazega}                    & Co-work                      & 71             & 717            \\
                                           & Friendship                   & 69             & 399            \\
                                           & Advice                       & 71             & 726            \\ \hline
\multirow{3}{*}{Vickers}                   & Get on                       & 29             & 240            \\
                                           & Best friends                 & 29             & 126            \\
                                           & Work                         & 29             & 152            \\ \hline
\end{tabular}
\caption{List of multilayer networks considered in this study. To facilitate visualization, the three biggest layers of each network have been selected. }
\label{tab:description}
\end{table}

\newpage

\section{The limitations of similarity-based techniques \label{sec:single_layer}}

The quality of the predictions is not only determined by the metrics used to compute the score, but also by the availability of information on its own, as pointed out by Jia et al. \cite{Jia2020}. For instance, in sparse networks, such as most real-world networks, most pairs of nodes will be without common neighbors and will be assigned exactly the same score, zero. In general, there will always be a set of scoreless links, limiting the maximum and minimum values of the AUC measure, to
\begin{equation}\label{eq:AUC_worst}
\text{AUC}_\text{min} = \frac{1}{2}(1+p_1)(1-p_2)  \qquad  \text{AUC}_\text{max} = \text{AUC}_\text{min} + p_1 p_2\,,
\end{equation}
where $p_1$ is the fraction of links with a score different from 0 among those links that will exist in the future, and $p_2$ is the same among the edges that will not exist.
Only in case of no scoreless links, that is when $p_1 = p_2 = 1$, we obtain $\text{AUC}_\text{min} = 0$ and $\text{AUC}_\text{max} =1$.

The above equation can be derived as follows. The limited amount of links present in a network bounds the prediction power of any similarity-based link prediction algorithm. This is due to the fact that scoreless links limit the value of AUC. The problem of having equal scores is not trivial and different statistical software packages choose different ways of solving the ties \cite{Muschelli2019Dec}. If we consider the classical approach of solving them (i.e., all links with the same score produce a single point in the TPR/FPR curve), the AUC is equivalent to the Mann Whitney Wilcoxon test \cite{Hanley1982Apr},
 so it reads:
\begin{equation}\label{eq:AUC_wilcoxon}
\text{AUC} = \frac{n'+0.5n''}{n}\,.
\end{equation}
The meaning of this expression is as follows. First, we pick two random links, one from the set of links that will exist in the future, which we denote as $P_1$, and one from the set of links that will not exist in the future, $P_2$. 
If the score of the link belonging to the first group is larger than the one from the second, $n'$ is incremented by 1.
If the score is the same, $n''$ is incremented instead. If this process is repeated $n$ times, equation \eqref{eq:AUC_wilcoxon} is equivalent to the classical area under the curve measured from the TPR/FPR curve \cite{Lu2011Mar}.

Following \cite{Jia2020}, the bounds on the AUC can be obtained by measuring the fraction of links in the set $P_1$ with a score different from 0, $p_1$, and similarly for the set $P_2$, yielding $p_2$. Hence, the fraction of links with a score equal to 0 in both sets will be $(1-p_1)$ and $(1-p_2)$, respectively. Thus, in the worst case scenario in which all links corresponding to $p_2$ have a score larger than $p_1$ we would have $n'/n=p_1(1-p_2)$ and $n''/n=(1-p_1)(1-p_2)$, so that
\begin{equation}\label{eq:AUC_worst}
\text{AUC}_\text{worst} = \frac{1}{2}(1+p_1)(1-p_2)\,.
\end{equation}
A similar argument allows us to determine the best possible AUC, i.e., the one in which all the links corresponding to $p_1$ have a score larger than the ones from $p_2$ (note that there will be a fraction $1-p_1$ of links that will exist in the future with score equal to 0 and hence below the ones corresponding to $p_2$) yielding
\begin{equation}\label{eq:AUC_best}
\text{AUC}_\text{best} = \text{AUC}_\text{worst} + p_1 p_2\,.
\end{equation}

\newpage

\section{Coefficients of the Multiplex Adamic-Adar metric}

Figures \ref{fig:coeff} and \ref{fig:coeff2} show the AUC and Precision of the MAA metric as a function of the coefficients $\eta_{x\alpha}$, for the five data sets not shown in the main text. 
These Figures are equivalent to Fig. 2 of the main text. 
Table \ref{tab:maximum} shows the combination of the coefficients ($\eta_{x1},\eta_{x2},\eta_{x3}$) that yields the maximum AUC and Precision for prediction of links in each layer, for all data sets considered. Note that such combination is also highlighted as a cross in all Figures. 
In some cases, the best results are achieved when some layers do not contribute at all ($\eta_{x\alpha} = 0$), while in others the contribution of all layers is important.
For instance, in the C. Elegans Genetic network, the prediction of new links in the physical association layer is mainly driven by the physical association layer ($\eta_{22}=0.87$ for the maximum AUC, in Table \ref{tab:maximum}).  However, the physical association layer plays an important role to predict direct interactions ($\eta_{12}=0.41$ for the maximum AUC, in Table \ref{tab:maximum}). 
Conversely, in the CS-Aarhus network, the contribution of the own layer where the links are being predicted is the most important one ($\eta_{xx} \simeq 1$ for all layers in Table \ref{tab:maximum}).

\begin{figure*}
\includegraphics[width=\linewidth]{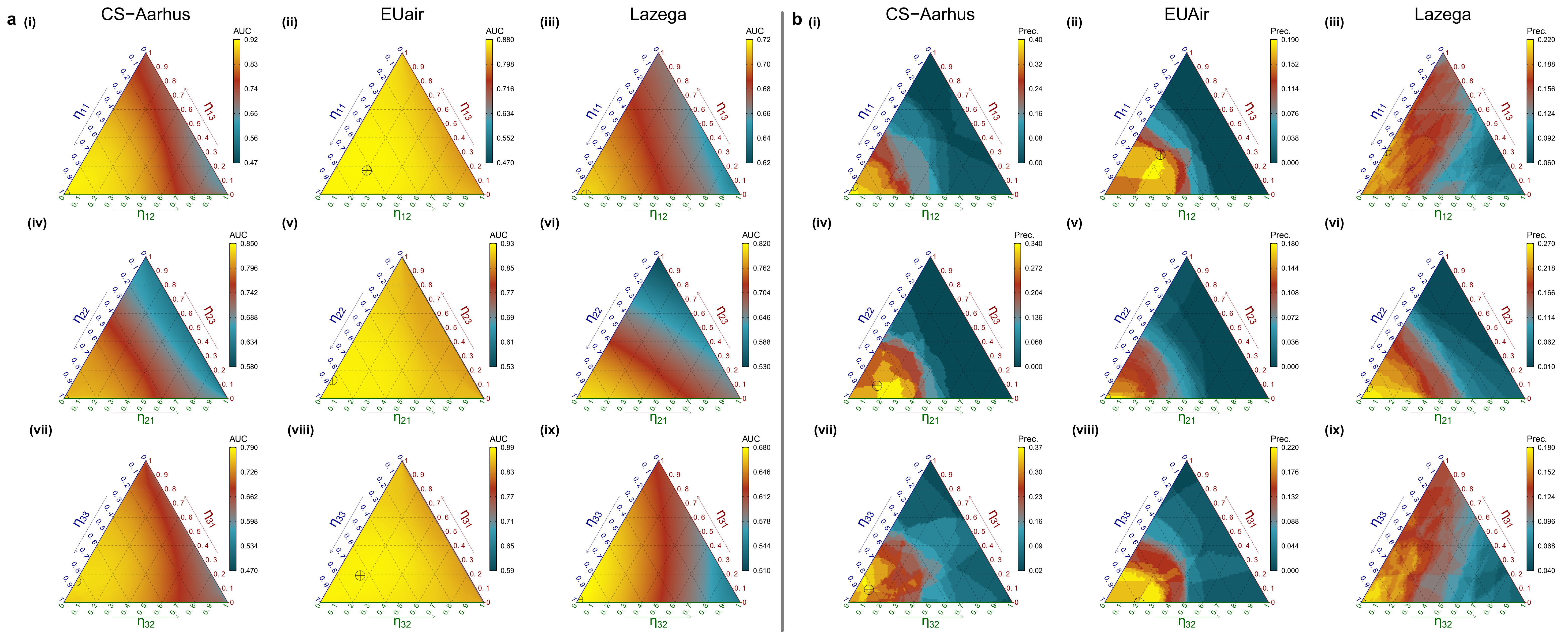}
\caption{\textbf{AUC (panel (a), left) and Precision (panel (b), right) of the MAA metric} for different values of the coefficients $\eta_{x \alpha}$,  indicating the weight of layer $\alpha$ in the prediction of new links in layer $x$.
Varying the values of two coefficients, the third is naturally fixed. 
Each column corresponds to a different data set, represented as a multiplex network of three layers. 
Each row corresponds to a prediction in a different layer $x$, see Table \ref{tab:maximum} for the corresponding interactions.
A cross indicates the maximum value for each plot, corresponding to the combination of coefficients ($\eta_{x1}$, $\eta_{x2}$, $\eta_{x3}$), reported in Table \ref{tab:maximum}, that maximizes AUC or Precision for the prediction of new links in layer $x$. 
\label{fig:coeff}}
\end{figure*}

\begin{figure*}
\includegraphics[width=\linewidth]{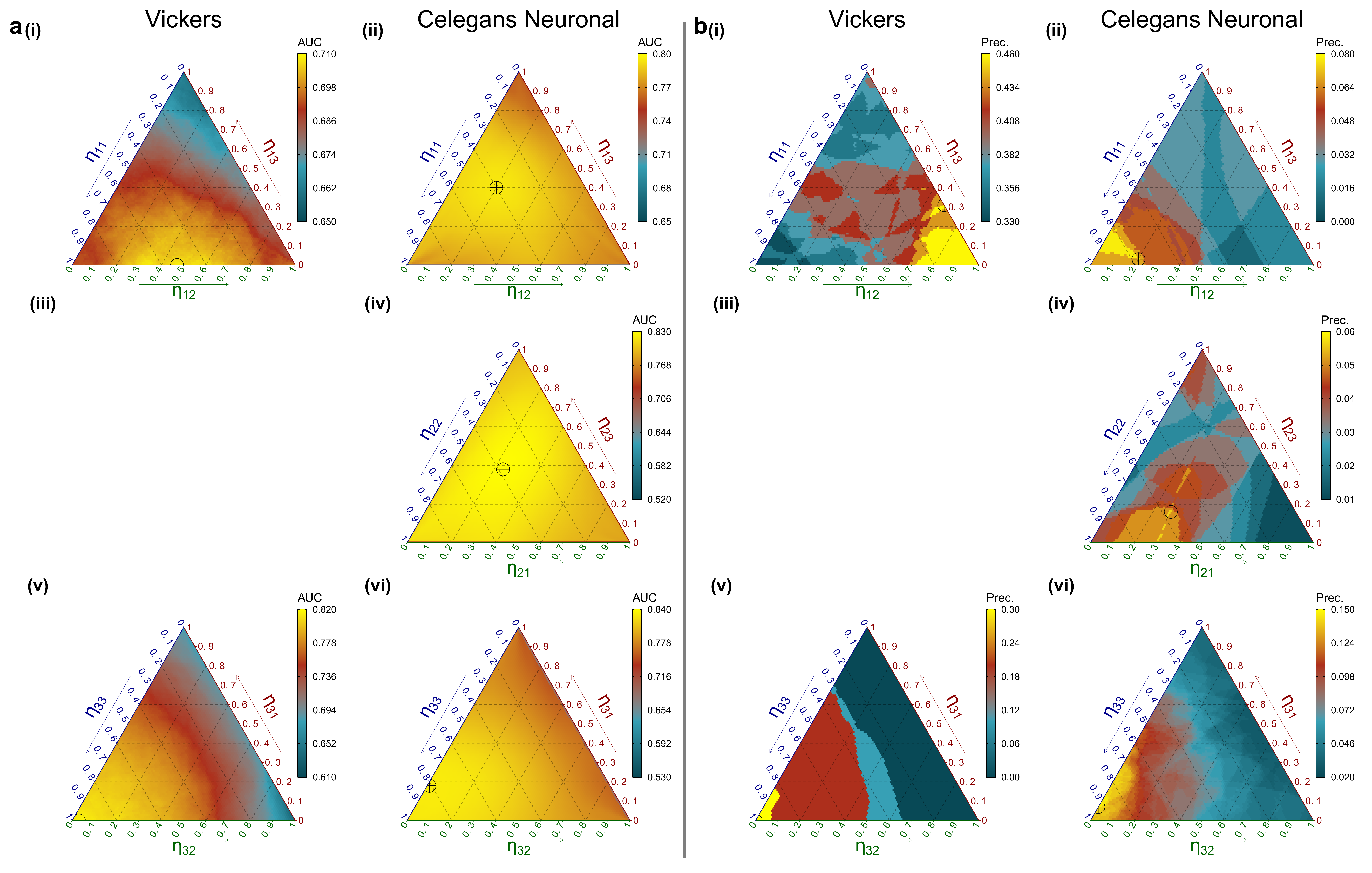}
\caption{\textbf{AUC (panel (a), left) and Precision (panel (b), right) of the MAA metric} for different values of the coefficients $\eta_{x \alpha}$,  indicating the weight of layer $\alpha$ in the prediction of new links in layer $x$.
Varying the values of two coefficients, the third is naturally fixed. 
Each column corresponds to a different data set, represented as a multiplex network of three layers. 
Each row corresponds to a prediction in a different layer $x$, see Table \ref{tab:maximum} for the corresponding interactions.
A cross indicates the maximum value for each plot, corresponding to the combination of coefficients ($\eta_{x1}$, $\eta_{x2}$, $\eta_{x3}$), reported in Table \ref{tab:maximum}, that maximizes AUC or Precision for the prediction of new links in layer $x$. 
The Vickers network is too small to have any non-overlapping links in the second layer. Therefore, links in such layers cannot be predicted.
\label{fig:coeff2}}
\end{figure*}

\begin{table}[]
\begin{tabular}{@{}llll@{}}
\toprule
\textbf{Network}                           & \textbf{Layer}               & \textbf{Maximum AUC} & \textbf{Maximum Precision} \\ \hline
\multirow{3}{*}{Copenhagen} & Phone calls                  & (0.29,0.4,0.31)      & (0.50,0.19,0.31)           \\
                                           & Facebook                     & (0.05,0.92,0.03)     & (0,0.98,0.02)              \\
                                           & SMS                          & (0.99,0.01,0)        & (0,0,1)                    \\ \hline
\multirow{3}{*}{C. Elegans genetic}        & Direct interaction           & (0.59,0.41,0)        & (0.61,0.22,0.17)           \\
                                           & Physical association         & (0.05,0.87,0.08)     & (0,0.83,0.17)              \\
                                           & Additive genetic interaction & (0.27,0.08,0.65)     & (0.02,0.39,0.59)           \\\hline
\multirow{3}{*}{C. Elegans neural}         & Electric                     & (0.40,0.20,0.40)     & (0.77,0.20,0.03)           \\
                                           & Chemical Monadic             & (0.24,0.38,0.38)     & (0.28,0.56,0.16)           \\
                                           & Chemical Plyadic             & (0.18,0.01,0.81)     & (0.07,0,0.93)              \\\hline
\multirow{3}{*}{CS-Aarhus}                 & Facebook                     & (0.99,0,0.01)        & (0.94,0,0.06)              \\
                                           & Leisure                      & (0,1,0)              & (0.13,0.78,0.09)           \\
                                           & Lunch                        & (0.15,0,0.85)        & (0.09,0.08,0.83)           \\\hline
\multirow{3}{*}{CKM Physicians}            & Advice                       & (0.61,0.27,0.12)     & (0.41,0.42,0.17)           \\
                                           & Discussion                   & (0.19,0.57,0.24)     & (0.39,0.08,0.53)           \\
                                           & Friendship                   & (0.01,0.28,0.71)     & (0,0.27,0.73)              \\ \hline
\multirow{3}{*}{EUair}                     & Airline1                     & (0.63,0.20,0.17)     & (0.52,0.20,0.28)           \\
                                           & Airline2                     & (0.01,0.86,0.13)     & (0,0.99,0.01)              \\
                                           & Airline3                     & (0.19,0.15,0.66)     & (0,0.21,0.79)              \\ \hline
\multirow{3}{*}{Lazega}                    & Co-work                      & (0.94,0.06,0)        & (0.69,0,0.31)              \\
                                           & Friendship                   & (0,1,0)              & (0,0.92,0.08)              \\
                                           & Advice                       & (0.02,0,0.98)        & (0,0,1)                    \\ \hline
\multirow{3}{*}{Vickers}                   & Get on                       & (0.53,0.47,0)        & (0,0.69,0.31)              \\
                                           & Best friends                 & -                    & -                          \\
                                           & Work                         & (0,0.03,0.97)        & (0,0,1)                    \\ \hline
\end{tabular}
\caption{Combination of coefficients ($\eta_{x1},\eta_{x2},\eta_{x3}$) that maximize AUC and Precision for the prediction of new links in each layer. The Vickers network is too small to have any non-overlapping links in the ``Best friends'' layer. Therefore, links in such layers cannot be predicted. }
\label{tab:maximum}
\end{table}

\newpage
\clearpage

\section{Comparison of MAA metric with other algorithms}

In Tables \ref{tab:comparison2} and \ref{tab:comparison3} we  compare  the  MAA  metric with other, state-of-the-art metrics for link prediction applied to the aggregated network representation, to predict links in the second and third layer of each data set, respectively. 
These Tables are equivalent to Table 2 of the main text, where we shown prediction of links in the first layer of each data set.

\begin{table}[h]
\begin{tabular}{@{}l|cccccccc@{}}
\toprule
\textbf{Method} & \textbf{CNS} & \textbf{CEG}  & \textbf{CEN}  & \textbf{CSA} & \textbf{CKM} & \textbf{EUA}  & \textbf{VIC}  & \textbf{LAZ}  \\ \colrule
AA    & 0.91           & \textbf{0.83} & 0.82          & 0.70          & 0.81           & 0.91          & -         & 0.60 \\
CN    & 0.91           & 0.82          & 0.80          & 0.71          & 0.80           & 0.91          & -          & 0.60 \\
JC    & 0.91           & 0.81          & 0.77          & 0.69          & 0.80           & 0.88          & -          & 0.60 \\
Katz  & \textbf{0.92}  & 0.79          & 0.79          & 0.67          & \textbf{0.90}  & 0.94          & -          & 0.49 \\
PA    & 0.77           & 0.79          & 0.70          & 0.66          & 0.57           & \textbf{0.95} & -          & 0.58 \\
MAA   & 0.91           & \textbf{0.83} & \textbf{0.83} & \textbf{0.84} & 0.81           & 0.92          & -          & \textbf{0.81} \\
\botrule
AA    & 0.32           & 0.00          & 0.03          & 0.00          & 0.11           & 0.04          & -          & 0.03 \\
CN    & 0.29           & 0.00          & 0.03          & 0.00          & 0.11           & 0.05          & -          & 0.03 \\
JC    & 0.24           & 0.00          & 0.03          & 0.08          & 0.09           & 0.00          & -          & 0.05 \\
Katz  & 0.25           & 0.00          & 0.00          & 0.00          & 0.00           & 0.00          & -          & 0.00 \\
PA    & 0.08           & 0.00          & 0.02          & 0.12          & 0.00           & 0.04          & -          & 0.04 \\
MAA   & \textbf{0.33}  & \textbf{0.26} & \textbf{0.06} & \textbf{0.33} & \textbf{0.13}  & \textbf{0.17} & -          & \textbf{0.26} \\
\botrule
\end{tabular}
\caption{\textbf{AUC (top) and Precision (bottom) to predict the second layer of each data set}, obtained for different metrics: Adamic Adar (AA), Common Neighbors (CN), Jaccard Similarity (JC), Katz distance (Katz), Preferential Attachment (PA), and Multiplex Adamic Adar (MAA). All metrics except for the MAA are applied to the aggregated network, including all information available. Best method is highlighted in bold. 
}
\label{tab:comparison2}
\end{table}

\begin{table}[h]
\begin{tabular}{@{}l|cccccccc@{}}
\toprule
\textbf{Method} & \textbf{CNS} & \textbf{CEG}  & \textbf{CEN}  & \textbf{CSA} & \textbf{CKM} & \textbf{EUA}  & \textbf{VIC}  & \textbf{LAZ}  \\ \colrule
AA    & 0.71          & 0.74          & \textbf{0.83} & 0.73          & 0.81           & 0.88          & 0.68         & 0.60 \\
CN    & 0.71          & 0.74          & 0.81          & 0.72          & 0.80           & 0.87          & 0.68          & 0.60 \\
JC    & 0.72          & 0.73          & 0.79          & 0.71          & 0.81           & 0.84          & 0.69          & 0.63 \\
Katz  & 0.65          & \textbf{0.75} & 0.82          & 0.69          & \textbf{0.90}  & 0.94          & 0.47          & 0.54 \\
PA    & 0.44          & \textbf{0.75} & 0.71          & 0.66          & 0.54           & \textbf{0.96} & 0.65          & 0.52 \\
MAA   & \textbf{0.72} & 0.74          & \textbf{0.83} & \textbf{0.78} & 0.81           & 0.88          & \textbf{0.82} & \textbf{0.68} \\
\botrule
AA    & 0.00           & 0.02          & 0.12          & 0.13          & 0.05           & 0.05          & 0.00          & 0.07 \\
CN    & 0.00           & 0.03          & 0.11          & 0.16          & 0.05           & 0.03          & 0.00          & 0.06 \\
JC    & 0.00           & 0.00          & 0.03          & 0.03          & 0.06           & 0.00          & 0.00          & 0.10 \\
Katz  & 0.00           & 0.00          & 0.00          & 0.00          & 0.00           & 0.00          & 0.00          & 0.00 \\
PA    & 0.00           & 0.02          & 0.08          & 0.11          & 0.00           & 0.03          & 0.00          & 0.08 \\
MAA   & \textbf{0.04}  & \textbf{0.04} & \textbf{0.15} & \textbf{0.37} & \textbf{0.08}  & \textbf{0.21} & \textbf{0.03} & \textbf{0.17} \\
\botrule
\end{tabular}
\caption{\textbf{AUC (top) and Precision (bottom) to predict the third layer of each data set}, obtained for different metrics: Adamic Adar (AA), Common Neighbors (CN), Jaccard Similarity (JC), Katz distance (Katz), Preferential Attachment (PA), and Multiplex Adamic Adar (MAA). All metrics except for the MAA are applied to the aggregated network, including all information available. Best method is highlighted in bold. 
}
\label{tab:comparison3}
\end{table}

\bibliographystyle{apsrev4-1}

\end{document}